IAC-25-A3,IP,151,x97770

# Rheological Lunar Regolith Simulants

Léonie E. Gasteiner[a]*, Alyona Glazyrina[a], Naomi Murdoch[a], Olfa D'Angelo[a]

[a] *Institut Superieur de l´Aeronautique et de l´Espace (ISAE-Supaero), Universite de Toulouse, Toulouse, France*
\* Corresponding Author: gasteinerleonie@gmail.com

**Abstract**

Regolith simulants are essential for space research and technology development. Yet, their physical properties often differ from those of true planetary soil, particularly when compared to regolith properties *in-situ*, that experience notably reduced gravity.

We focus on lunar regolith simulants and explore various techniques to modify existing simulants to replicate the mechanical/rheological behavior of Moon regolith in true lunar conditions. Our results are validated against data from in-situ tests conducted during the Luna and Apollo missions, enabling a direct comparison of physical properties of our enhanced simulants to true Lunar regolith, on the Moon. Analyzed in the Mohr-Coulomb model framework, the friction angle of most simulants is generally close to that of true regolith, but the measured cohesion is often higher on the Moon, notably due to the reduced gravity environment. We propose a method to increase the cohesion of an existing simulant and assess the mechanical behavior of our rheological regolith simulant using a standardized geotechnical, shear test. The experimental results are then directly compared to in-situ data, providing a quantitative basis for evaluating the fidelity of the enhanced simulants.

**Nomenclature**
ρ     Bulk density (g/cm$^3$)
φ     Mohr Coulomb angle of internal friction (degrees)
c     Mohr Coulomb cohesion (kPa)¨
Bo    Granular Bond number (-)

**Acronyms/Abbreviations**
European Space Agency (ESA)
*In-Situ* Resource Utilization (ISRU)
Polydimethylsiloxane (PDMS)

## 1. Introduction

As humans prepare to return to the Moon, understanding the lunar environment becomes a priority to ensure safety of astronauts and equipment. The behavior of the lunar soil has been a challenge for past lunar missions [1]. While the 20th century Moon missions provided invaluable insight about the lunar environment, many aspects of its surface behavior remain unresolved [2, 3].

Lunar regolith is the layer of dust covering the Moon's surface, formed over millions of years by the fragmentation of volcanic rocks due to meteorite impacts. Its chemical composition is now known rather precisely [4]. However, its mechanical properties and flow behavior (rheology) remain complex to predict. Only a relatively small amount of lunar regolith could be brought back to Earth, which has limited the scope of experimental studies. To overcome this limitation, lunar regolith simulants, materials mimicking certain aspects of regolith, have been developed and are now essential for testing technologies and *In-Situ* Resource Utilization (ISRU) research.

While current simulants faithfully reproduce specific physical and chemical characteristics of real lunar regolith, significant differences remain when considering their mechanical response and flow behavior. Many factors influence the flow behavior of granular materials, including gravity [5, 6, 7] which makes *in-situ* data comparison essential when planning missions to the lunar surface [8]. A rheological lunar regolith, a material that would replicate the flow behavior and mechanical properties of true lunar soil as it is found on the Moon would support the development and testing of technologies such as rover wheels, lunar infrastructures and shelters, or lander footpads.

The aim of this work is to characterize and modify lunar regolith simulants to more accurately replicate the mechanical properties of true lunar soil, as found on the Moon. We first describe the lunar regolith simulant chosen for the experiments, and the methodology used for its characterization and modification. The results obtained are compared to *in-situ* measurements from past lunar missions.

## 2. Methods
### 2.1 The granular Bond number
The mechanics of granular materials are complex, and the relationship between the microscopic forces acting between grains and the bulk behavior of the material is still not fully understood [9]. These microscopic forces control the flow behavior of the material by influencing how easily particles separate, rearrange, and resist movement [10, 11].

The flow behavior of powders can be characterized using the granular Bond number, a dimensionless quantity defined by the ratio of intrinsic forces, notably interparticle cohesive forces, to extrinsic forces, the weight of the particles [12, 13].

$$Bo = \frac{F_{int}}{F_{ext}}$$

The granular Bond number measures the balance between cohesive forces within the material and the gravitational force acting on it; explaining why granular materials exhibit different flow behaviors under different gravitational fields. Following this line of reasoning, for a simulant to mimic the flow behavior of lunar regolith on the Moon, the higher gravity on Earth must be compensated by an increase in intrinsic cohesive forces.

*2.2 Mohr-Coulomb Failure Framework*
The stress-strain data obtained from direct shear tests, analyzed within the Mohr-Coulomb failure framework, describes the point at which a material fails under shear and normal stress [14]. This approach is chosen for two main reasons: it is widely used for analyzing granular materials, especially in geotechnics, and it was used for interpreting the mechanical data collected on lunar regolith during the Apollo missions [4].

Within the Mohr-Coulomb framework, two parameters describe a granular material: cohesion, c, and angle of internal friction φ. Using this method allows direct comparison between our results and the *in-situ* measurements obtained during past lunar exploration campaigns.

It is important to distinguish between interparticle cohesive forces and cohesion as a Mohr-Coulomb parameter. While both describe the tendency of grains to remain bound together, they represent different physical quantities.

*2.3 Characterization of the EAC-1A simulant*
EAC-1A is a lunar regolith simulant developed by the European Astronaut Center (EAC), situated in Cologne, Germany, for the European Lunar Exploration Laboratory (LUNA), a lunar analog facility. This simulant offers a promising baseline for testing due to its availability to European researchers. Few publications so far have characterized this material [15], and its mechanical properties have not yet been comprehensively measured. EAC-1A is classified as a silty sand; its grain size distribution lays within the upper and lower Apollo boundary, with a median at ~70 μm. The cohesion of EAC-1A was estimated by Engelschion et. al. [15] to be 0.38 kPa by lowering a plastic plate into a box filled with the material. The composition of the simulant is close to that of true lunar regolith, with the exception of higher alkali minerals content and the presence of minerals such as quartz and chlorite not found or rarely found on the Moon.

We determine the mechanical properties of the EAC-1A lunar regolith simulant using direct shear tests at a shearing speed of 1 mm/min and under normal stresses of 1, 5, and 10 kPa. The resulting cohesion and angle of internal friction are determined using the Mohr–Coulomb framework. Each measurement is repeated 3 times and the standard deviation was determined for each parameter.

*2.4 Modification of the EAC-1A simulant*
The objective of the modification is to increase the cohesive interparticle forces within the simulant to values representative of true lunar regolith under low gravity conditions. To achieve this, the particles of the simulant are coated with a thin layer of a silicone-based polymer, polydimethylsiloxane (PDMS), known for its adhesive and cohesive properties.

This method is inspired by the work of Gans et. al. [16], who demonstrated that coating monodisperse spherical glass beads with PDMS polymer significantly increased interparticle cohesion. The coating process involves mixing a small amount of boric acid, PDMS, and water into the simulant using a heated kitchen mixer, until the material reached a smooth, powdery texture. The full procedure is described in details in ref [16].

Two different batches of modified simulant are prepared, with estimated polymer coating thicknesses of 20 nm and 50 nm (labeled EAC-1A_CC20 and EAC-1A_CC50, respectively). Note that this estimate is based on theoretical calculation for spherical particles, which is not the case for lunar regolith simulants. Both modified samples are characterized using the same direct shear testing procedure as the unmodified simulant, a constant shearing speed of 1 mm/min and under normal stresses of 1 kPa, 5 kPa, and 10 kPa, allowing for direct comparison of the results. Each measurement is repeated 3 times and standard deviation is computed for each parameter.

The three different batches of simulant, EAC-1A, EAC-1A_CC20, and EAC-1A_CC50 are observed under a microscope to get a better understanding of how the PDMS coating is getting distributed in the material and how it may affect the interparticle cohesive forces between the grains.

## 3. Results & Discussion

Direct shear tests are performed on the unmodified EAC-1A simulant, as well as on the two batches of modified simulants. Table 1 summarizes the resulting cohesion and angle of internal friction determined using the Mohr-Coulomb failure criterion. They are graphically represented in Figure 1.

Table 1. Results of cohesion and angle of internal friction

| Material | Bulk Density (g/cm$^3$) | Cohesion (kPa) | Angle of internal friction (degree) |
| --- | --- | --- | --- |
| EAC-1A | 1.45 | 1.42±0.45 | 39.44±2.14 |
| EAC-1A_CC20 | 2.17 | 2.26±0.86 | 35.15±4.67 |
| EAC-1A_CC50 | 2.04 | 1.71±0.47 | 38.83±2.28 |

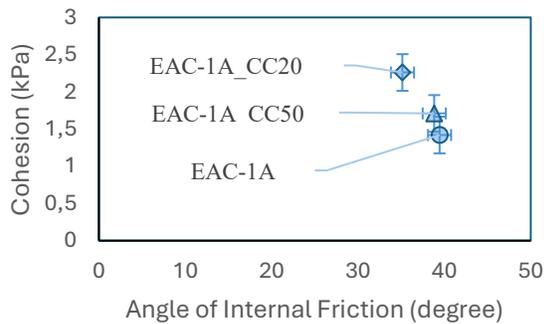

Figure 1. Mohr-Coulomb parameters, cohesion and angle of internal friction, of the materials tested via direct shear, and average values for true lunar regolith *in-situ*.

### 3.1 Characterization of EAC-1A

The characterization of the unmodified EAC-1A simulant by direct shear yields a cohesion of 1.42 ± 0.45 kPa and an angle of internal friction of 39.44 ± 2.14 degrees. The only previously reported characterization of EAC-1A was by Engelschion et. al. [15]; they report a much lower cohesion of c=0.38 kPa. Engelschion et al. do not report an angle of internal friction for EAC-1A. Such discrepancies are explained by the different method used to find the material's cohesion: a plastic plate was lowered into a simulant-filled box, instead of the more classical direct shear we employ here.

### 3.2 Modification of EAC-1A

As shown in Table 1 and Figure 1, the modified simulant EAC-1A_CC20 exhibited a significant increase in cohesion, about + 60% from the unmodified material EAC-1A. This is the expected outcome, consistent with the adhesive properties of the PDMS polymer and with the objectives of the modification.

However, the modified sample EAC-1A_CC50 displays a lower cohesion than EAC-1A_CC20, although the amount of polymer coating should lead to a superior level of cohesion. This is unexpected, as Gans et. al. [16] reported a monotonic increase in cohesion with the amount of coating added to the material (corresponding to coating thickness for spherical glass beads).

Several factors can explain this trend. First, the EAC-1A particles have high surface roughness and irregular shapes, which may prevent the formation of a uniform PDMS layer. Instead, the polymer might accumulate in surface crevices or on some of the particles, resulting in uneven coating distribution and localized variations in adhesion. Second, at higher thicknesses, the PDMS layer may begin to behave as a lubricant rather than a binder, reducing interparticle friction and thereby lowering the apparent cohesion. This is particularly true for shear tests, where normal force is imposed.

For both modified materials, changes in the angle of internal friction are relatively small, in the order of 1% to 10%, suggesting that the PDMS primarily influences cohesive behavior.

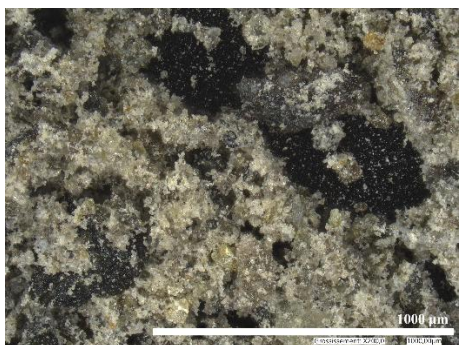
(a) EAC-1A_CC20

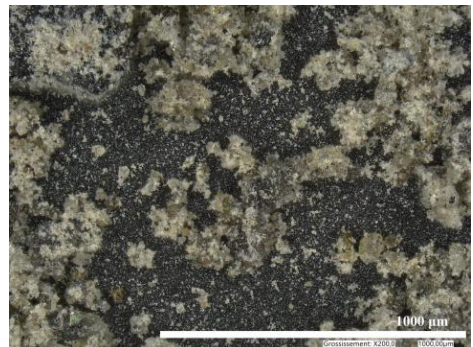
(b) EAC-1A_CC50

Figure 2. Images of the EAC-1A_CC20 (a) and EAC-1A_CC50 (b) samples observed under an optical microscope at magnification 200

To verify the hypotheses formulated above, the samples are observed under an optical microscope. Figure 2 shows the microscopies of the different samples observed with magnification 200.

This observation allows to see the effect of the PDMS coating on the organization and repartition of the grains of the material. In all materials, very small particles tend to stick on larger sand grains. However, in the EAC-1A_CC50 (fig. 2b), many smaller particles are visible around the larger grains. These small particles, instead of adhering to the large grains, form clusters, held together by the PDMS. Instead of homogeneously coating all the EAC-1A particles, the polymer seems to affect predominantly the smallest particles, effectively modifying the particle size distribution of the material.

## 4. Conclusion

In this study, we investigated the mechanical properties of the EAC-1A lunar regolith simulant and explored a method to increase its interparticle cohesive forces to be more representative of true lunar regolith under low-gravity conditions. The unmodified EAC-1A as well as the modified materials were characterized using direct shear testing. The resulting Mohr–Coulomb parameters were determined and compared to previously reported values for EAC-1A [15] and for true lunar regolith as measured on the Moon during the Apollo missions [4].

Cohesion values measured were almost 4 times higher than those reported by Engelschion et. al. [15]; likely due to differences in measurement methodology. The simulant was modified by coating its grains with thin layers of PDMS polymer, in quantities equivalent to thicknesses of approximately 20 nm and 50 nm on spherical glass beads. The coating of EAC-1A_CC20 produced the desired increase in cohesion, confirming the potential of PDMS to enhance interparticle cohesion. However, the testing of EAC-1A_CC50 yielded lower cohesion than expected, possibly due to uneven coating distribution. In both cases, the angle of internal friction only had small variations, indicating that the modification primarily affected cohesion rather than frictional interactions. Observing the coated simulant with an optical microscope confirmed these hypotheses, adding that the wide grain size distribution of the material may affect the influence of the coating.

These results highlight the potential of targeted surface modifications to improve the mechanical fidelity of lunar regolith simulants. Further cohesion-enhancing methods will be tested in the future to refine simulants ability to reproduce the mechanical response of true lunar regolith. Achieving accurate cohesion values is particularly important for reproducing lunar flow behavior on Earth, where higher gravity can be compensated by increased intrinsic interparticle cohesive forces. This approach will contribute to the development of more representative simulants, supporting technology developments and ISRU research, essential for sustainable lunar exploration.

## Acknowledgements

The authors thank Alexia Duchene for her assistance with the direct shear testing equipment and microscopies, and Jeanne Bigot for providing the base code used in data processing. L.G. acknowledges the ESA ISEB scholarship for funding her participation to IAC 2025. O. D'A. acknowledges financial support from the French National Center for Space Studies (CNES) under fellowship 24-357. N.M. acknowledges funding support from the French Space Agency (CNES) and from the European Research Council (ERC) GRAVITE project (Grant Agreement N° 1087060).